\documentclass[12pt]{elsarticle}

\usepackage{amssymb} 
\usepackage{graphicx}
\usepackage{dcolumn}
\usepackage{bm}

\journal{Physics Letters B}

\begin{document}

\begin{frontmatter}

\title{Testing the black disk limit in $pp$ collisions at very high energy }%

\author{P. Brogueira}
\ead{pedro.brogueira@ist.utl.pt}
\address{Departamento de F\' \i sica, IST, Av. Rovisco Pais, 1049-001 Lisboa, Portugal }
\author{J. Dias de Deus}
\ead{jorge.dias.de.deus@ist.utl.pt}
\address{CENTRA, Departamento de F\' \i sica, IST, Av. Rovisco Pais, 1049-001 Lisboa, Portugal}

\date{\today}%

\begin{abstract}
We use geometric scaling invariant quantities to measure the approach, or not, of the imaginary and real parts of the elastic scattering amplitude, to the black disk limit, in $pp$ collisions at very high energy.
\end{abstract}


\end{frontmatter}

Saturation phenomena are expected  to dominate QCD physics at high energy and high matter density \cite{ref1,ref2}. At LHC and at ultra high energy cosmic rays these effects should become easily detected. One of the possible consequences of such physics is the setting up of the black disk regime in the description of soft physics, namely total cross-section, elastic cross-section, elastic differential cross-section, etc., as discussed for instance in \cite{ref3,ref4} (see also \cite{ref5}).

Non linear differential equations include, in a natural way, saturation effects. This happens with the well known logistic equation, which can be seen as a simplified version of the B-K equation \cite{ref6}. For a general discussion on evolution and saturation, in the present context see \cite{ref7} and \cite{ref8}. In this paper we apply the logistic equation to the evolution of the imaginary part of the impact parameter elastic amplitude, ${\rm Im} G(s,b^2)$.

In \cite{ref9} we have argued that the evolution with the energy, $\sqrt{s}$, of ${\rm Im} G(s,b^2)$ - or of the Fourier-Bessel transform, ${\rm Im} F(s,t)$ - qualitatively describe the evolution of the differential elastic cross-section, $d \sigma / d t$, in particular in the small $|t |$ region. Here we present quantitative tests for the approach of ${\rm Im} F(s,t)$ to the black disk in the region $|t |\lesssim |t_0 |$, $t_0$ being the position of the first diffractive zero. As most of the cross-sections are concentrated in the small $|t|$ region, any test for the approach to the black disk has, at least, to work at small $|t|$.

We study next the evolution of the profile function $\Gamma(s,b^2)$, or the imaginary part of the impact parameter elastic amplitude, 
\begin{equation}
\label{equa1}
\Gamma(s,b^2) \equiv {\rm Im} G(s,b^2),
\end{equation}
and we consider the logistic equation
\begin{equation}
\label{equa2}
\frac{\partial \Gamma}{\partial b}= -\frac{1}{\gamma}(\Gamma-\Gamma^2),
\end{equation}
where $\gamma >0$ is a positive constant, and $b \simeq \frac{2}{\sqrt{s}}\ell$, $\ell$ being the angular momentum and $\sqrt{s}$ the center of mass energy. As, from unitarity,
\begin{equation}
\label{equa3}
0 \leq \Gamma \leq 1,
\end{equation}
$\partial \Gamma / \partial b \leq 0$, which means that $\Gamma$ is a decreasing function of $b$, becoming eventually constant at small $b$. This means that saturation occurs first at small $b$. At large $b$, $\Gamma$ decreases exponentially $(\Gamma \sim \exp (-b / \gamma))$. The parameter $\gamma$, which we take as a constant, controls the long range behaviour of the strong forces and can be associated to the two pion exchange diagram. A solution of (\ref{equa2}) - not the most general one - is
\begin{equation}
\label{equa4}
\Gamma(b^2,s)=\frac{1}{\exp{\frac{b-R}{\gamma}+1}},
\end{equation}
$R$ being a positive quantity, $R>0$. In comparisons with data, for $\sqrt{s}\gtrsim 50 \; {\rm GeV}$, we shall fix $\gamma$, $\gamma=1.1 \; {\rm mb}^{1/2}$, which means that the dependence of $\Gamma$ on $\sqrt{s}$ is exclusively contained in $R \rightarrow R(s)$. One sees that $R$ is a radial scale parameter, in fact the only relevant parameter in the black disk limit. It is then clear, from (\ref{equa4}), that there is also an evolution equation in $R$, or in the energy, 
\begin{equation}
\label{equa5}
\frac{\partial \Gamma}{\partial R}= \frac{1}{\gamma}(\Gamma-\Gamma^2).
\end{equation}
One further sees that $\partial \Gamma/ \partial R \geq 0$. As $\sigma_{tot.}(s)$, defined as
\begin{equation}
\label{equa6}
\sigma_{tot.}(s)=2 \pi \int \Gamma(b^2,s) db^2
\end{equation}
experimentally increases with $\sqrt{s}$, for $\sqrt{s} \gtrsim 20\; {\rm GeV}$, and $\Gamma$, at least for some values of $b$, increases with $R$, it is clear that $R(s)$ increases with energy. In the $R (s) \to \infty$ limit we obtain the black disk.

Regarding the elastic cross-section, 
\begin{equation}
\label{equa7}
\sigma_{el.}(s)=\pi \int \left[ |\Gamma(b^2,s)|^2+|{\rm Re}F(t,s)|^2 \right] db^2,
\end{equation}
and neglecting the ${\rm Re} F(s,t)$ contribution, one sees that, necessarily, because of (\ref{equa3}), in the evolution to a black disk, the ratio $\sigma_{el.}/\sigma_{tot.}$ has to increase, eventually reaching the value $1/2$. In fact the observed increase of the ratio $\sigma_{el.}/\sigma_{tot.}$ at the CERN/SPS \cite{ref10} was the first clear indication of the possibility for the evolution towards a black disk to occur.

In general, in the calculation of $d \sigma / d t$, 
\begin{equation}
\label{equa8}
\frac{d \sigma}{d t}=\frac{\sigma_{tot.}^2}{16 \pi} \left[ \frac{{\rm Im}F(t,s)}{{\rm Im}F(0,s)} \right]^2 \left( 1+\rho^2(t,s) \right).
\end{equation}
where
\begin{equation}
\label{equa9}
\rho(t,s) \equiv \frac{{\rm Re}F(t,s)}{{\rm Im}F(t,s)},
\end{equation}
one needs to know the real part of the amplitude. As we consider energies above $50 \; {\rm GeV}$ we shall assume that $\sigma^{tot}_{pp} \simeq \sigma^{tot.}_{\bar{p}p}$ and write a derivative dispersion relation \cite{ref11} in the form
\begin{equation}
\label{equa10}
\frac{{\rm Re}F(t,s)}{s}= \left[ \frac{\pi}{2} \frac{\partial}{\partial \ln s} \right] \frac{{\rm Im}F(t,s)}{s},
\end{equation}
to estimate $\rho(t,s)$, \cite{ref9}.

The black disk regime is obtained from (\ref{equa4}) by taking the limit
\begin{equation}
\label{equa11}
\frac{\gamma}{R} \rightarrow 0,
\end{equation}
or, as $\gamma$ is constant,
\begin{equation}
\label{equa12}
R \rightarrow \infty ,
\end{equation}
or
\begin{equation}
\label{equa13}
\sigma^{tot.} \rightarrow \infty.
\end{equation}
In the limits (\ref{equa12}) or (\ref{equa13})$\Gamma(b^2,s)$ satisfies geometric scaling \cite{ref12},
\begin{equation}
\label{equa14}
\Gamma(b^2,s) \, \, \,  \raisebox{-5pt}{$\overrightarrow{\scriptstyle R(s) \rightarrow \infty}$} \, \, \, \varphi \left( b / R \right).
\end{equation}
Working in the $(t,s)$ plane, defining the scaling variable
\begin{equation}
\label{equa15}
\tau \equiv -t \sigma^{tot.},
\end{equation}
one obtains for the imaginary and for the real part \cite{ref9} of $F(s,t)$:
\begin{eqnarray}
{\rm Im} F(s,t) & = & {\rm Im} F(s,0) \varphi(\tau),\label{equa16}\\
{\rm Re} F(s,t) & = & {\rm Re} F(s,0) \left. \left. \frac{d}{d \tau} \right[ \tau \varphi(\tau) \right]. \label{equa17}
\end{eqnarray}

In the particular case of the black disk, with interest for us here, using the variable
\begin{equation}
\label{equa18}
x=\sqrt{\frac{\tau}{2 \pi}}
\end{equation}
we obtain for the imaginary part, 
\begin{equation}
\label{equa19}
\varphi(\tau) =\frac{2}{x} J_1(x),
\end{equation}
and, for the real part,
\begin{equation}
\label{equa20}
\left. \left. \frac{d}{d \tau} \right[ \tau \varphi(\tau) \right]=J_0(x).
\end{equation}
It should be noticed that as $\rho(s,0)$ asymptotically tends to zero in the scaling limit only the imaginary part matters.

Our strategy in testing the possible approach to the black disk, as $\sqrt{s}$ increases, is to find quantities that are geometric scaling invariants, as functions of the scaling variable $\tau=-t \sigma^{tot.}$, and ask the questions: are these quantities approaching scaling values as the energy and $\sigma^{tot.}$ increase?

\begin{figure}[t]
\begin{center}
\includegraphics[width= 9.5 cm]{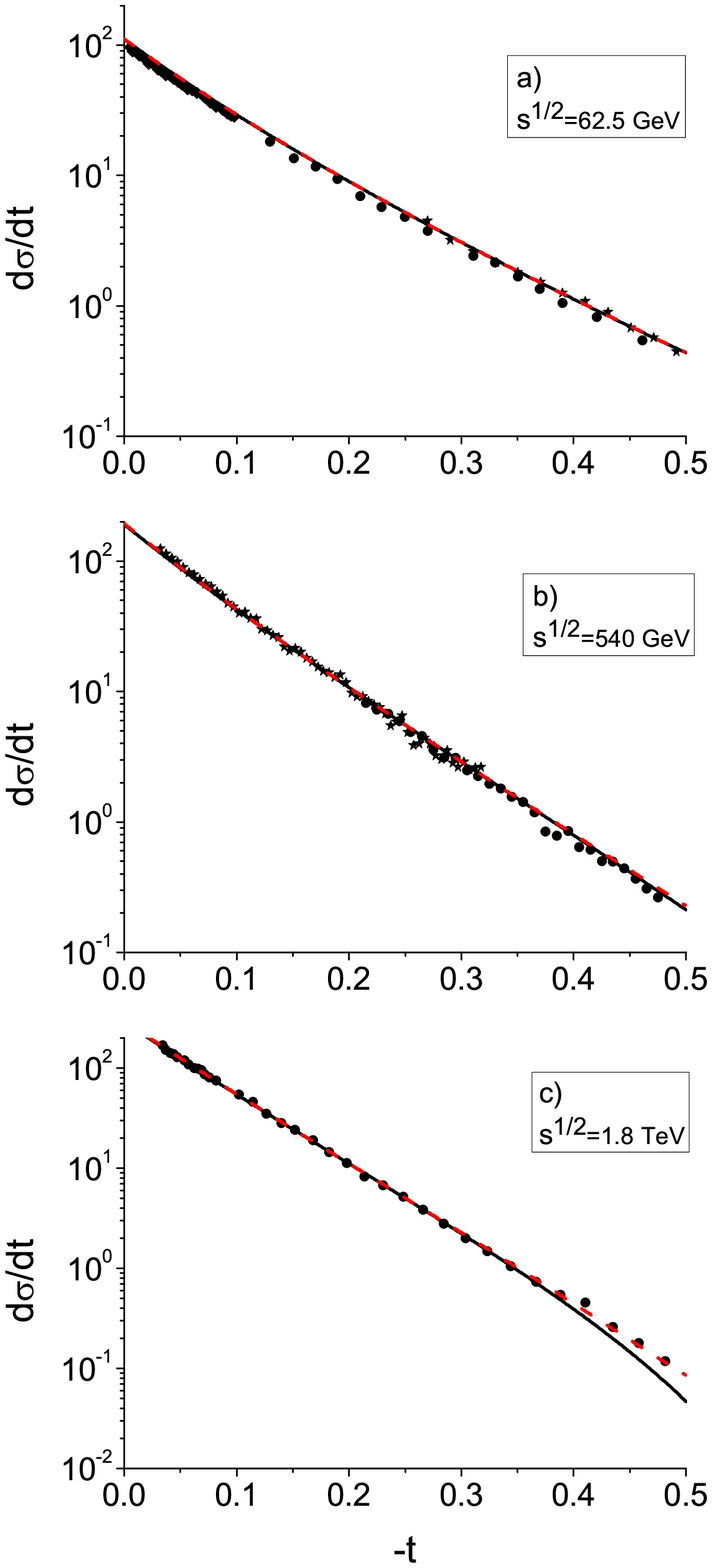} 
\end{center}
\vskip -1.5 true cm  
\caption{\label{fig1} Differential elastic cross-section $d\sigma /dt$ for $pp$ collisions at high energy. Data from \cite{ref12,ref13,ref14}. Curves obtained from (\ref{equa4}), for the imaginary part contribution (full Lines) with $\gamma=1.1 \; {\rm mb}^{1/2}$, and  (\ref{equa10}) for real part contribution (dashed lines for imaginary part plus real part contribution).}
\end{figure}

In Figs. \ref{fig1}.a), b) and c) we present the comparisons of our model, (\ref{equa4}) and (\ref{equa10}),  for $d \sigma / d t$ with low $|t|$ precision data at ISR, $62.5 \; {\rm GeV}$, \cite{ref12}, SP$\bar{\rm P}$S, $540 \; {\rm GeV}$, \cite{ref13} and Tevatron, $1.8 \; {\rm TeV}$, \cite{ref14}. The parameter $\gamma$ was fixed, the only parameter left being $R^2(s)$. In the three cases the values found for $\sigma^{tot.}(s)$ agree with experiment \cite{ref15}.

\begin{figure}[t]
\begin{center}
\hskip -0.9 true cm
\includegraphics[width= 9.5 cm]{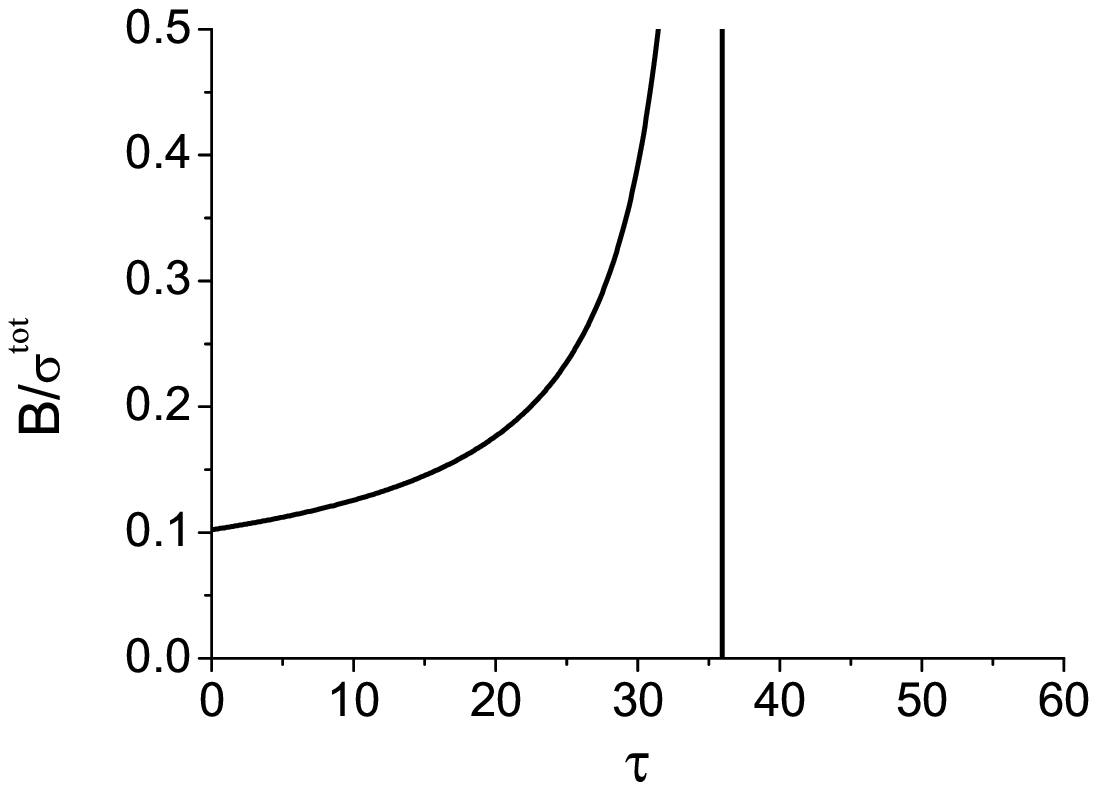}      
\end{center}
\vskip -1.0 true cm 
\caption{\label{fig2} The ratio $B(s/t) / \sigma^{tot.} (s)$, (\ref{equa21}), as a function of the scaling variable $\tau$, (\ref{equa15}), in the case of the black disk at asymptotic energy.}
\end{figure}

We shall next construct the quantity
\begin{equation}
\label{equa21}
\frac{B(t,s)}{\sigma^{tot.}(s)}\equiv \frac{1}{\sigma^{tot.}(s)} \frac{\partial}{\partial t} \left( \ln \frac{d \sigma}{d t}\right),
\end{equation}
where $B(t,s)$ is the local slope parameter. In the asymptotic black disk limit (\ref{equa21}) is a function of the scaling variable $\tau$, see Fig. \ref{fig2}.
We further have:
\begin{eqnarray}
i) \; \frac{B(t \rightarrow 0,s)}{\sigma^{tot.}(s)}&=&\frac{1}{8 \pi} \;\; \; ({\rm natural \; \; units}) \label{equa22}\\
&=&0.102 \; {\rm GeV}^{-2}{\rm mb}^{-1} ; \nonumber \\
ii) \; \; \; \; \;  \frac{B(t,s)}{\sigma^{tot.}(s)}\; \; \; &=&0 \;\;  {\rm for} \; \; \tau=35.9201 \; {\rm GeV}^2{\rm mb} \label{equa23},
\end{eqnarray}
the position of the first zero of $J_1(x): \; x=3.8317$;\\

\hskip 1.9 true cm {\it iii)} The ${\rm Re}F(s,t)$ contribution to $B(t,s)/ \sigma^{tot.}(s)$ (not shown in figure \ref{fig2}) approaches zero at the zero of $J_1(x)$: at the zero of $J_1(x)$, $J_0(x)$ is a maximum or a minimum.

\begin{figure}[t]
\begin{center}
\hskip -0.9 true cm
\includegraphics[width= 9.5 cm]{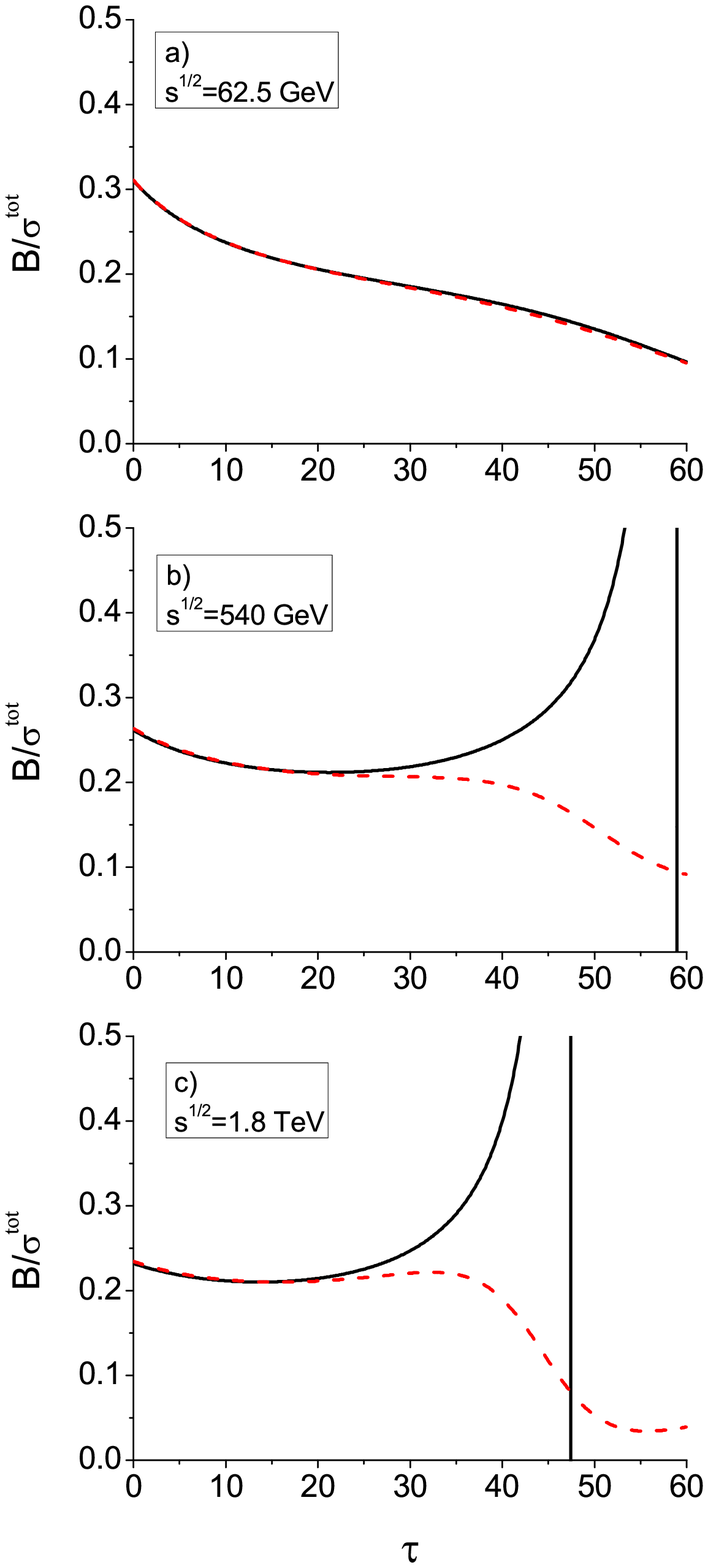}      
\end{center}
\vskip -1.0 true cm
\caption{\label{fig3} The ratio $B(s/t) / \sigma^{tot.} (s)$, (\ref{equa21}), as a function of $\tau$, (\ref{equa15}), at different $pp$ energies. The full lines represent real imaginary part contributions, dashed lines the real part contribution alone. At $\sqrt{s}=62$ GeV there is no evidence for an asymptotic black disk.}
\end{figure}

In Fig. \ref{fig3} a), b) and c) we show the quantity (\ref{equa21}) in the cases of previously studied situations, Fig. \ref{fig1}. We note that at ISR, $\sqrt{s}=62.5 \; {\rm GeV}$, there is not yet evidence for the first zero of ${\rm Im} F(s,t)$, and $B / \sigma^{tot.}$, at small $|t|$, decreases as $|t|$ increases. At $\sqrt{s}~=540 \; {\rm GeV}$ and $1.8 \; {\rm TeV}$ the zero is there and $B / \sigma^{tot.}$ becomes flatter at small $|t|$ (exponential in $|t|$ behaviour) increasing at large $|t|$ (see Fig. \ref{fig2}).

\begin{figure}[t]
\begin{center}
\hskip -0.9 true cm
\includegraphics[width= 9.5 cm]{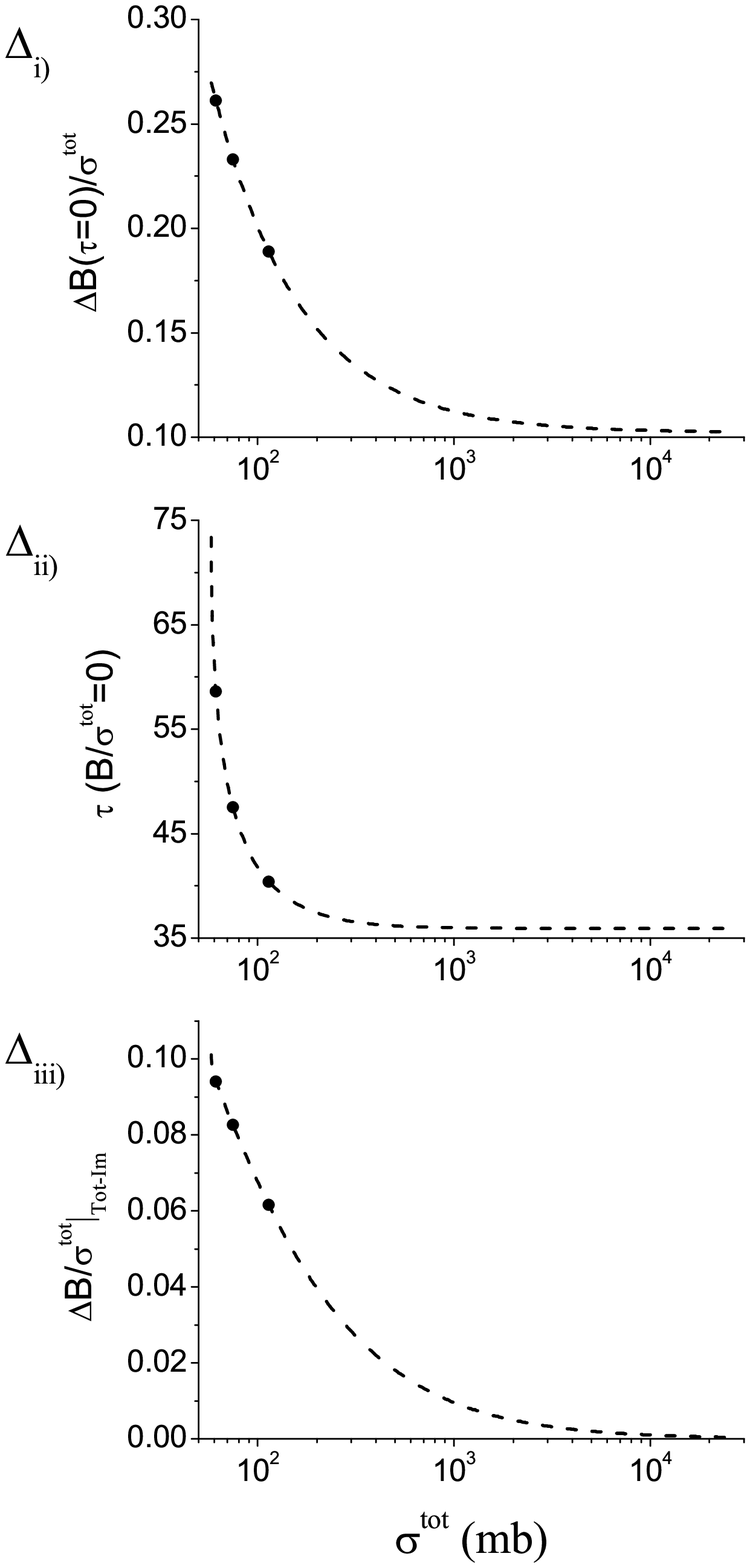}      
\end{center}
\vskip -1.0 true cm
\caption{\label{fig4} The approach to the black disk measured by the differences $\Delta_{i)}, \Delta_{ii)}$ and $\Delta_{iii)}$, Eqs. (\ref{equa24}) to (\ref{equa26}), as a function of $\sigma{tot.}(s)$. The points and the dashed lines correspond to our model. We have included our expectation for LHC energy, 14 TeV, $(\sigma^{tot.} (s) \simeq 110 \; mb)$.}
\end{figure}

In order to describe the approach to the black disk in our model, let us introduce, in connection with remarks {\it i)},{\it ii)} and {\it iii)} above, quantities measuring the approach to the black disk (BD) as function of $\sigma$ (or energy):
\begin{eqnarray}
\Delta_{i)} & \equiv & \frac{B(t=0)}{\sigma^{tot.}}- \left. \frac{B(t=0)}{\sigma^{tot.}} \right|_{BD}, \label{equa24}\\
&& \nonumber\\
\Delta_{ii)} & \equiv & (-t \sigma^{tot.} )- (-t \sigma^{tot.} )|_{BD} \; \; \; \; \; {\rm at \; \; the \; \; zero \; \; of \; \; Im}F(s,t),  \label{equa25}\\
&&\nonumber\\
\Delta_{iii)} & \equiv & \left. \frac{B}{\sigma^{tot.}} \right|_{\rm Real+Im.}- \left. \frac{B}{\sigma^{tot.}} \right|_{\rm Im.} \; \; \; \; \; {\rm at \; \; the \; \; zero \; \; of \; \; Im}F(s,t),  \label{equa26}
\end{eqnarray}
In Fig. \ref{fig4} {\it i)}, {\it ii)} and {\it iii)} we show $\Delta_{i)}$, $\Delta_{ii)}$ and $\Delta_{iii)}$ as function of $\sigma^{tot}$, respectively, including our expectation for LHC, 14 TeV, ($\sigma^{tot} \simeq 110 \; {\rm mb}$).

It should be pointed out that the points in Fig. \ref{fig4}, except the LHC point, come from experiment, but were obtained with a model satisfying asymptotic black disk behavior. More independent analysis should be attempted. Tests of the approach to the black disk limit, in particular $\Delta_{i)}$, (\ref{equa24}), have been previously discussed (see, for instance,   \cite{ref5} and \cite{ref17}).

~\\
{\bf Acknowledgments}\\
~\\
We would like to thank Carlos Pajares and Jaime Muniz for discussions.\\

\end{document}